\documentclass[preprint,12pt]{iopart}

\usepackage{graphicx}
\usepackage{iopams}
\expandafter\let\csname equation*\endcsname\relax
 \expandafter\let\csname endequation*\endcsname\relax
\usepackage{amsmath}
\usepackage{url}
\usepackage{enumerate}

\newcommand{\CC}{\ensuremath{\mathbb{C}}}
\newcommand{\ZZ}{\ensuremath{\mathbb{Z}}}

\begin{document}

\title{An Accumulative Model for Quantum Theories }

\author{Chris Thron}
\address{$^1$Department of Mathematics, Texas A\& M University - Central Texas, 1001 Leadership Place,
 Killeen, TX 76549, United States}
\ead{thron@tamuct.edu}

\begin{abstract}
For a general quantum theory that is describable by a path integral formalism, we propose a physical model that depicts the spacetime universe as a ``slice'' or ``snapshot'' of an accumulative process that occurs within a higher-dimensional space. We give a rigorous mathematical characterization of the model, and show that it give predictions that are nearly identical to the given quantum theory.   The new model  is neither local nor causal  in spacetime, but is both local and causal  in the higher-dimensional space in which spacetime resides. The probabilistic nature of the squared wavefunction is a natural consequence of the model. We verify the model
 with simulations, and discuss discrepancies from conventional quantum theory that might be detectable via experiment. Finally, we discuss some physical implications of the model. 
\medskip

\end{abstract}
\maketitle
\section{Introduction}
\label{sec:intro}
In a previous paper \cite{ThronWatts},  a model of quantum particle transmission was proposed to explain some puzzling aspects of the quantum theory.  According to the model, particle detection is the outcome of a signal accumulation process which occurs in an extra, non-spacetime dimension (which we refer to as the $a$-dimension). The complex wavefunction corresponds to in-phase and quadrature-phase components of an amplitude and phase-modulated carrier signal field that is present throughout spacetime augmented by the $a$-dimension. We postulated that the location of particle detection is determined when an accumulated signal reaches a threshold, and proved that the Born probability rule is a mathematical consequence.  In this way, we have exhibited a deterministic (but statistically random) process whose outcomes obey the same probability distribution as predicted by quantum theory.   However, the paper gives no explanation of the origin or formation of the carrier signal field required for the model.

 The current paper provides a more comprehensive interpretation of quantum probabilities by taking a related but somewhat different approach. The approach is based on the observation that both quantum mechanics and quantum field theory may be derived from a  path integral formalism. We conjecture that path integrals correspond to a universal physical process which effectively performs a numerical integration. As in the previous paper, this process unfolds in a non-spacetime dimension, and the observable universe is the outcome of the process upon attaining a threshold.

The paper is organized as follows.  Section~\ref{sec:presentation} presents a simplified preliminary mathematical model which illustrates the basic model structure. We demonstrates the model's ability to generate quantum probabilities both theoretically and with simulations.  Section~\ref{sec:adjusted} gives a more detailed model which is designed to conform  more closely with the hypothesized physical processes involved. Section~\ref{sec:exper} discusses the possibility of experimental verification of the model; and Section~\ref{sec:disc} gives a summary discussion.

\section{Preliminary model}
\label{sec:presentation}
Let $\mathcal{U}$ represent the space of all possible states of the observable universe. We emphasize that any $u \in \mathcal{U}$ expresses the entire state of the universe over all times, not just its state at a single time. We do not need to specify whether we are employing a quantum-mechanical or field-theoretic representation of these states -- our argument does not depend on the specific nature of $\mathcal{U}$.

In both quantum-mechanical or the field-theoretic representations of $\mathcal{U}$, the wavefunction can be expressed in terms of a path integral $\Psi:\mathcal{U} \rightarrow \CC$ of the form::
\begin{equation}\label{eq:PI}
\Psi(u) \equiv  \sum_{ \gamma \in \Gamma_u} e^{iS(\gamma)},
\end{equation}
where $\Gamma_u$ is a space of paths corresponding to the state $u$, and $S(\gamma)$ is the action associated with the path $\gamma$. Here we have used a sum rather than an integral to facilitate the connection with simulations that we will describe later.
 Here each path $\gamma$ corresponds to a unique $u$, so $\{\Gamma_u\}_{u \in \mathcal{U}}$ are disjoint. We assume that the number of paths corresponding to each state is the same, so that $|\Gamma_u|$ is independent of $u$.

The path integral is associated with a probability distribution:
\begin{equation}\label{eq:prob}
P_S(u) \equiv \frac{| \Psi(u) |^2}{\sum_{v\in\mathcal{U}} | \Psi(v) |^2}.
\end{equation} 
The fact that this probability is written in terms of summations qua integrals suggests that some sort of accumulative process could be involved. The main purpose of this paper is to show that such an interpretation is indeed feasible, and provides a simple, plausible explanation of the hidden dynamics that give rise to quantum theories.  

We define an accumulation process as follows. Given the sequence of paths  $\gamma_1, \gamma_2, \ldots$ in $\Gamma$,  we define an  \emph{accumulated amplitude} $\mathcal{A}_K~(K \in \ZZ_+)$ as:
\begin{align}\label{eq:acc}
\mathcal{A}_K \equiv \Sigma_{k=1}^{K} e^{iS(\gamma_k)}.
\end{align}
One possible interpretation of each factor $e^{iS(\gamma_k)}$ is as the phasor representation \cite{Phasors} of an oscillation (of unknown frequency) whose phase is given by $S(\gamma_k)$\footnote{This  phase is measured with respect to an arbitrarily-defined reference--this arbitrariness produces $U(1)$ gauge invariance.}. The summation then corresponds to the complex amplitude of a harmonic oscillator (with the same frequency)  that is successively perturbed by these oscillations. 

\begin{figure}[h]
\begin{center}
\includegraphics[width=5.5in]{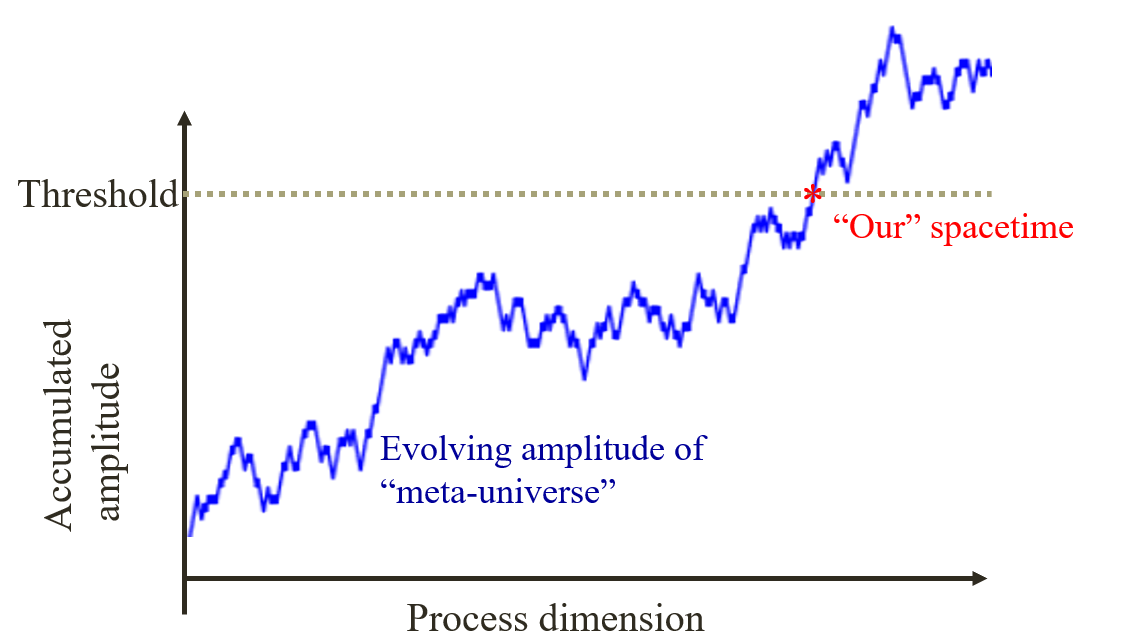}
\end{center}
\caption{ Evolution of the ``meta-universe'', which produces spacetime when its state attains a certain threshold. \label{fig:process}}
\end{figure}

Although we are using discrete notation, the sequence $\{\gamma_k\}$ should be thought of as a discrete approximation of a path-valued function of a continuous index, corresponding to a continuously-varying path within the space $\Gamma$  of all possible paths. This variation is presumed to  be governed by some stochastic process which  uniformly samples $\Gamma$ over the long term.  As $\gamma_k$ varies, the corresponding (potential) universe state $u_k \equiv u_{\gamma_k}$ also varies. In the process we will define, the accumulated amplitude  grows until it passes a predefined threshold at a particular index $K$, at which point $u_K$ gives the state of the spacetime universe that we experience. Thus the accumulated amplitude can be considered as indicating the ``level'' of the process, and the observable universe   represents the state of the process once it has attained a particular level.  On a simple level, we may envision this process as  a ``meta-universe'' that evolves in some nonobservable, non-spacetime dimension until the accumulated amplitude hits a certain level,  and the meta-universe is revealed as ``our'' spacetime (Figure~\ref{fig:process}). As a visual analogy, consider a giant soap bubble that  contorts as it floats through the air (for a marvellous video of this, see \cite{Day}). Over time the water in the membrane evaporates, until ithe remaining water content is no longer sufficient to maintain the necessary surface tension \cite{Troutner}.  A hole forms, and the bubble disintegrates as though ``rolled up''. The shape of the bubble at the moment of disintegration depends on the entire history of the entire bubble. Although it appears that receding edge of the disintegrating bubble is a ``traveling object''  (just as we  observe ``objects'' that travel through space and time), in fact the moving edge simply traces over the shape of the membrane that  was already there. (Naturally there are limits to this analogy--the meta-universal process we are postulating is \emph{not} a process in time, and we are not necessarily implying that the universe is ``disintegrating''.) 

\begin{figure}[h]
\begin{center}
\includegraphics[width=5.5in]{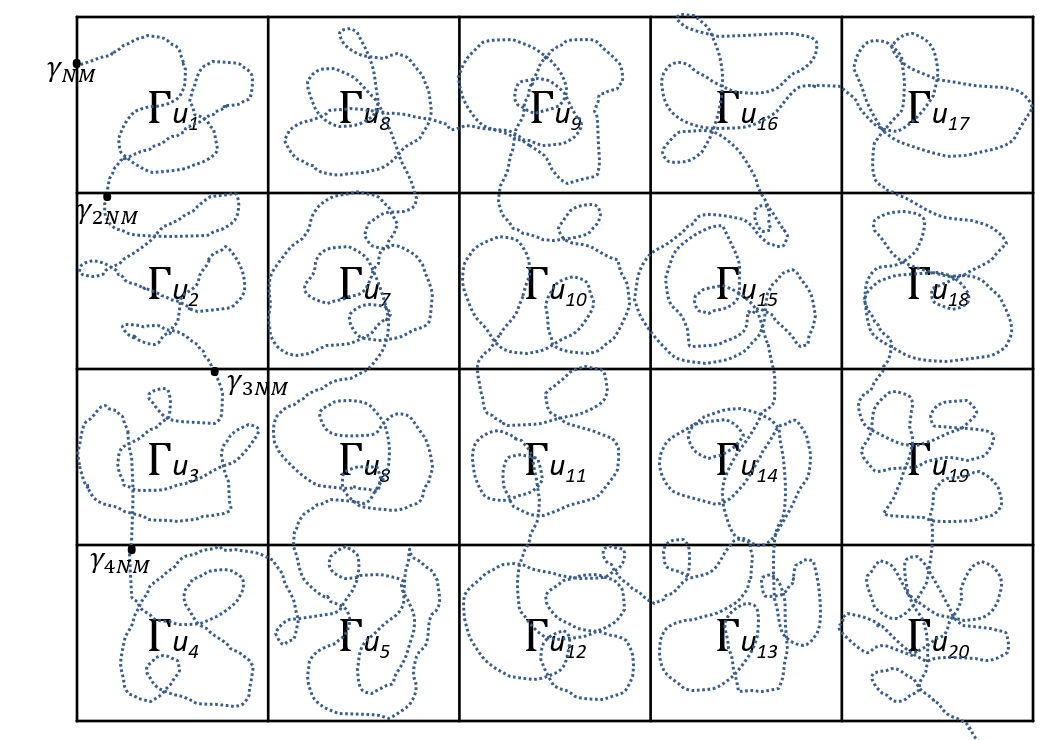}
\end{center}
\caption{ Schematic illustration of preliminary model of a quantum system. The dotted line represents the trajectory of paths $\gamma_k$ $(k=0,1,2,\ldots)$ as it wanders through the space of all paths. Boxes label regions within path space that are associated to different physical states $u_j$. \label{fig:trajectorySimple}}
\end{figure}

By imposing a few simple conditions on the sequences $\{ \gamma_k \}$ and $\{u_k\} $ in our proposed meta-universal stochastic process, we may recover the state probabilities \eqref{eq:prob}. These conditions are  as follows:    
\begin{enumerate}[(a)]
\item\label{cond1}
There exists $N \gg 1$ and $M \gg 1$ such that  $u_{kNM} = u_{kNM+1} =  \ldots =u_{(k+1)NM-1} ,  \forall k \in \ZZ_{\ge 0}$;
\item\label{cond2}
For each $k \in \ZZ_{\ge 0}$, the sequence $\{ \gamma_{kN}, \gamma_{kN+1}, \ldots, \gamma_{(k+1)N-1}\} $ uniformly samples $\Gamma_{u_{kN}}$;
\item\label{cond3}
The sequence $\{u_{NM}, u_{2NM}, \ldots \}$ 
uniformly samples $\mathcal{U}$.
\end{enumerate}
Figure~\ref{fig:trajectorySimple} schematically illustrates the system described by conditions (a)-(c). The figure shows the trajectory $\{\gamma_k\}, k=0,1,2,\ldots$ as it passes through path space $\Gamma$. Each box represents the set of paths corresponding to a physical state $u$, The trajectory takes $NM$ steps in each box before passing to the next, and these steps represent a uniform sample of $\Gamma_u$ (in our subsequent model, these rigid assumption will be relaxed). The significance of $M$ will be explained later.

Let $\eta_k  ~(k=1,2, \ldots)$ be a sequence of independent, identically distributed (i.i.d.) complex-valued  random variables with zero mean and finite variance, and define:
\begin{equation}\label{eq:Apdef}
\mathcal{A}'_K = \sum_{k=1}^K \eta_{\lceil k/N \rceil } e^{iS(\gamma_k)}.
\end{equation}

Finally, given $\Theta > 0$, we define the \emph{threshold index} as the random variable:
\begin{equation}\label{eq:Apdef2}
K_{\Theta} \equiv \min( k | |\mathcal{A}'_k| < \Theta \text{ and } |\mathcal{A}'_{k+1}| \ge \Theta).
\end{equation}
Given the above conditions and definitions, we have the following result:
\medskip

\noindent
{\bf Proposition}: As $\Theta \rightarrow \infty$, we have for every element $u^* \in \mathcal{U}$ 
\begin{equation}
P( u_{K_{\Theta}} =u^*)  \xrightarrow[\Theta \rightarrow \infty]{} P_S(u^*).
\end{equation}\label{eq:bigProp}
In other words, the probability distribution on $\mathcal{U}$ at the stopping time corresponding to attaining the threshold $\Theta $ agrees with the probability distribution  (\ref{eq:prob}) obtained from the path-integral formalism, for sufficiently large values of $\Theta$. In other words, the process described produces a probability distribution on states that approaches the distribution obtained from the conventional path integral formalism of quantum mechanics.

The proof of this proposition is similar to that given in \cite{ThronWatts}, and the reader is referred to that paper for more details.  First we write
\begin{equation}
\Theta = N\sqrt{M}\theta,
\end{equation}
where $N, M,$ and $\theta$ can all be taken as arbitrarily large.  then \eqref{eq:Apdef} can be rewritten as
\begin{equation}
\frac{\mathcal{A}'_{KN}}{\Theta} = \frac{\mathcal{A}'_{KN}}{\theta N \sqrt{M}} =\frac{1}{\theta N\sqrt{M}} \sum_{k=1}^K  \eta_{k}\left(\sum_{n=1}^N  e^{iS(\gamma_{(k-1)N+n})} \right).\label{eq:simeqfull} 
\end{equation}
The sum in parentheses resembles \eqref{eq:PI}, except it is a sum over $N$ terms instead of over all elements of $\Gamma_u$.  If $N$ is large and the terms approximate a uniform random sample, then the sum will approximate the value of $\Psi$. This gives us:
\begin{equation}
\frac{\mathcal{A}'_{KN}}{\Theta} \underset{N \rightarrow \infty} \longrightarrow    \frac{ 1}{\theta|\Gamma_u|\sqrt{M}}\sum_{k=1}^K  \eta_{k} \Psi(u_{\lceil k/M \rceil}). \label{eq:simeq}
\end{equation}
By breaking the sum in \eqref{eq:simeq} into sections of $M$ terms each, we obtain (given that $K$ is a multiple of $M$):
\begin{equation}
\frac{\mathcal{A}'_{KN}}{\Theta} \underset{N \rightarrow \infty} \longrightarrow    \left(\frac{1}{\theta|\Gamma_u|}\right) \sum_{\ell=1}^{ K/M} \Psi(u_{\ell}) \nu_{\ell}, 
\label{eq:first}
 \end{equation}
 where the $\nu_{\ell}$ are i.i.d. Gaussian random variables defined by:
 \begin{equation}
 \nu_{\ell} \equiv \frac{1}{\sqrt{M}} \sum_{m=1}^{M} \eta_{(\ell-1)M+m}.  
 \end{equation}
The right-hand side of \eqref{eq:first} represents a random walk in $\mathbb{C}$, which approaches a Brownian motion as $\theta \rightarrow \infty$.  The value of  $K_{\Theta}$ represents the value of $K$ for which the Brownian motion hits the boundary $|z|=1$ in an absorbing random walk. As explained in \cite{ThronWatts}, near the boundary the probability density of an absorbing Brownian motion is proportional  (to first order) to the distance from the boundary. This can be used to show that for any $k$, the probability $P( K_{\theta N \sqrt{M}} = k | u_k = u^*)$ is approximately  proportional to $E[ | \eta_k \Psi(u^*) |^2]$, which is  proportional to $|\Psi(u^*)|^2$ (see \cite{ThronWatts} for details). Also, $P( u_k = u^*)$ is independent of $u^*$ when $1 \ll k < K_{\theta N \sqrt{M}}$, so
\begin{align}
 & P( K_{\Theta} = k ~\cap ~u_{K_{\theta N \sqrt{M}}} = u^*) = P( K_{\theta N \sqrt{M}} = K | u_K = u^*)P( u_K = u^*) \\
 \implies & P( K_{\Theta} = k ~\cap ~u_{K_{\Theta}} = u^*) \propto |\Psi(u^*)|^2, 
 \end{align}\label{eq:final}
  and summing \eqref{eq:final} over $k$ gives the desired result \eqref{eq:bigProp}.

\begin{figure}[h]
\begin{center}
\includegraphics[width=6.5in]{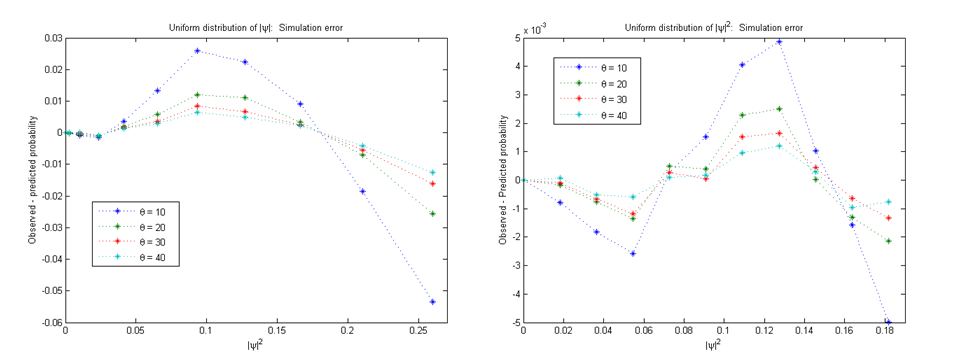}
\end{center}
\caption{Deviations of computed probabilities  from quantum values, for simulated preliminary accumulation model with $\theta = 10, 20, 30, 40$ and $M=10000$, where $\{\eta_k\}$ are i.i.d. standard normal random variables. Each simulation was run 100,000 times. All simulations used 11 configurations $u$. For the figure on the left, the $|\psi(u_j)| \propto j,~(j=0,\ldots, 10)$, while for the figure on the right, $|\psi(u_j)|^2 \propto j~(j=0,\ldots, 10)$.  \label{fig:sim2}}
\end{figure}

Figure ~\ref{fig:sim2} shows the 
results of simulations of the model specified by conditions (\ref{cond1})--(\ref{cond3}) and equations (\ref{eq:Apdef})--(\ref{eq:Apdef2}). The simulations were performed on a discrete system with 11 possible states. To shorten computational time, the simulation was based on equation (\ref{eq:simeq}) rather than performing the full computation (\ref{eq:simeqfull}) on a path-by-path basis. The random variables $\{u_{NM},u_{2NM},\ldots \}$  referred to in (\ref{cond3}) were generated uniformly randomly.   The curves show the difference between the simulated probabilities and actual probabilities for two different probability distributions $|\Psi|^2$, for different values of the threshold $\theta$.  The errors are shown on the $y$-axis, versus the actual probability values which are shown on the $x$-axis.  As $\theta$ increases, the errors decrease: for $\theta = 40$, the maximum error is under 5 percent. The pattern of error apparently depends on the type of probability distribution being modeled. However, in both cases the larger probabilities are underestimated, and there is a range of intermediate probabilities that are overestimated. These phenomena may possibly enable an experimental test of the model--see Section~\ref{sec:exper} for further discussion of this issue.

\section{Refined model}\label{sec:adjusted}
The model described in Section~\ref{sec:presentation} is based on the simple idea that the universe results from a `'meta-universe'' performing a random walk within a space of possible universes, which is ``snap-shotted'' at one point in the process. But it has some awkwardly artificial features:
\begin{itemize}
\item
Why should $\{u_k\}$ remain constant for intervals of size $MN$?
\item
What is the  physical significance of the $\eta_k$'s?
\end{itemize}
As to the first point, instead of supposing that $\{u_k\}$ remains constant on intervals of size $N$, we may suppose that $\{u_k\}$ varies slowly with $k$, so that
\begin{equation}
\textrm{Pr}(u_{k+1} \neq u_k) = \mathcal{O}\left(\frac{1}{N}\right).
\end{equation}
Supposing that $\{\gamma_k\}_{k=1.2.3.\ldots}$ is generated by a Markov process, it is reasonable to suppose that residence times in each $u$ state visited are (approximately) i.i.d. geometrical random variables. This is because under reasonable conditions,  hitting times in Markov chains are asymptotically exponentially distributed \cite{Aldous}. (The geometrical distribution is the discrete analog of the exponential distribution.) Accordingly, we may modify the model by replacing the constant $M$ with a geometrically-distributed random variable with the same mean.  

As to the second point, we must recognize that we have failed to account for the fact that in practice we never measure the state of the entire universe, but only a subsystem. So we must take into account the effect of variations in the external system during the accumulation process. Accordingly we let $\Omega$ be the possible states of the measured subsystem, while $\Omega'$ denote the possible states of the universe external to the measured subsystem. Thus we may represent any element $u \in \mathcal{U}$ uniquely as 
$u = (w, w')$, where $w \in \Omega$ and $w' \in \Omega'$. 

We suppose that any path in $\Gamma$ can be factored into a part for $\Omega$ and a part for $\Omega'$: more precisely, that there are path spaces $\Gamma_{\Omega}$ and $\Gamma'_{\Omega '}$ respectively such that any $\gamma \in \Gamma$ can be decomposed as $\gamma = (g, g')$ where $g \in \Gamma_{\Omega }, g' \in \Gamma'_{\Omega `}$, and such that  $u_{\gamma} = (w_g,w'_{g'})$. We define   $\Gamma_w \equiv \{g | w_g = w\}$, and suppose (as in the simple model)  that
$|\Gamma_w|$ is independent of  $w \in \Omega$. We similarly define $\Gamma'_{w'}$, with $|\Gamma'_{w'}|$ independent of $w'$. Finally, we suppose that the action $S$ is additive: $S(\gamma) = S((g,g') = S(g)+S(g')$.  From this it follows that we may write:
\begin{equation}
\Psi(u) = \Psi((w,w')) = \psi(w)\phi(w'),
\end{equation}
 where 
\begin{equation}
\psi(w) \equiv  \sum_{ g \in \Gamma_w} e^{iS(g)}; \quad \phi(w') \equiv  \sum_{ g' \in \Gamma'_{w'}} e^{iS(g')}.
\end{equation}
 
We may also rewrite (\ref{eq:acc}) as
\begin{align}\label{eq:acc2}
\mathcal{A}_K \equiv \Sigma_{k=1}^{K} e^{iS(g_k)}e^{iS(g'_k)}.
\end{align}
We now postulate the existence of a Markov chain $\{(g_1,g'_1), (g_2,g'_2), \ldots \}$ that describes the evolution of the meta-universe. Define inductively a sequence of random times $\{ \mathcal{X}_k \}$ such that  
\[\mathcal{X}_0 \equiv 1	; \qquad \mathcal{X}_{k+1} \equiv \min( j>\mathcal{X}_{k} | w'_j \neq w'_{\mathcal{X}_{k}}). \] 
We suppose the Markov chain has transition probabilities such that  the external state $w'_j$ varies more rapidly than the observed state $w_j$. This is a reasonable assumption since the external state is vastly larger and has many more possibilities for variation.  So to close approximation we can say that $w_j$ only changes when $w'_j$ changes and $w_j$ is constant on each interval $[ \mathcal{X}_{k}, \mathcal{X}_{k+1})$. We may also define inductively a sequence of random times 
$\{ \mathcal{Z}_k \}$ such that 
\[ \mathcal{Z}_0 = 1; \qquad  \mathcal{Z}_{k+1} \equiv \min( j > \mathcal{Z}_{k} | w_{\mathcal{X}_j}  \neq w_{\mathcal{X}_{Z_k}}).\] 
Finally, we suppose that the paths vary much faster than the states, so that  the space $
\Gamma_{w_{\mathcal{X}_k},w'_{\mathcal{X}_k}}\equiv
\Gamma_{w_{\mathcal{X}_k}}\times \Gamma'_{w'_{\mathcal{X}_k}}$ is uniformly sampled on the time interval $[\mathcal{X}_k, \mathcal{X}_{k+1}-1]$. 

\begin{figure}[h]
\begin{center}
\includegraphics[width=5.5in]{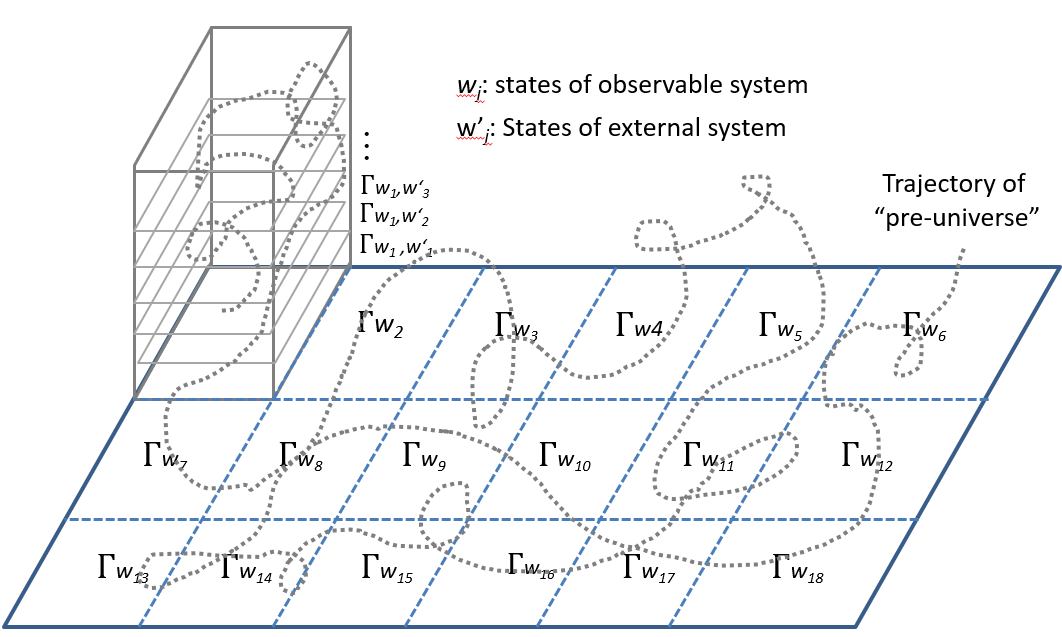}
\end{center}
\caption{ Refined model of a quantum system. The dotted line and squares represents the trajectory  $(\gamma_1,\gamma_2,\ldots )$ and observed physical states respectively. The squares $\Gamma_{w_k}$ have a further substructure, which is shown explicitly for square $\Gamma_{w_1}$. The detailed structure of square $\Gamma_{w_1}$ is represented as a rectangular box  partitioned into subsets, where each subset gives paths associated with the same observed state $w_1$ but different unobserved states $w'_j$ .  The trajectory uniformly samples each rectangular subset uniformly before passing on to the next subset, and the box above    \label{fig:trajectoryRefined}}
\end{figure}

The Markov chain described above  is depicted schematically in Figure~\ref{fig:trajectoryRefined}.
As the figure shows, each set $\Gamma_{w_j}$ is further subdivided into subsets indexed by $(w_j,w'_k)$, where $w_j$  and  $w'_k$ represent respectively the mesured and unmeasured portions of the universe's state.  Within each subset  $\Gamma_{w_j,w'_k}$, the process wanders uniformly just as in the squares in Figure~\ref{fig:trajectorySimple}.

The model may be formulated mathematically as follows:
\begin{enumerate}[(A)]
\item\label{pcond1a}
There a sequence $\{\mathcal{X}_1 \le  \mathcal{X}_2 \le  \ldots\}$ such that $\{\mathcal{X}_{K+1} - \mathcal{X}_K\}$ are  i.i.d. geometrically-distributed random variables with expected value $N \gg 1$ and  $w'_{\mathcal{X}_K} = w'_{\mathcal{X}_K+2} =  \ldots =w'_{\mathcal{X}_{K+1}-1} ~  \forall K \in \ZZ_{\ge 0}$.
\item\label{pcond1b}
There a sequence $\{\mathcal{Z}_1 \le  \mathcal{Z}_2 \le  \ldots\}$ such that $\{\mathcal{Z}_{K+1} - \mathcal{Z}_K\}$ are  i.i.d. geometrically-distributed random variables with expected value $M \gg 1$ and  $w_{\mathcal{X}_{\mathcal{Z}_K}} = w_{\mathcal{X}_{\mathcal{Z}_K + 1}} =  \ldots =w_{\mathcal{X}_{\mathcal{Z}_{K+1} - 1}} ~  \forall K \in \ZZ_{\ge 0}$.
\item\label{pcond2a}
For each $K \in \ZZ_{\ge 0}$, the sequences $\{(g_{\mathcal{X}_K} g'_{\mathcal{X}_K}),  \ldots, (g_{\mathcal{X}_{K+1}-1}, g'_{\mathcal{X}_{K+1}-1}) $ uniformly sample $\Gamma_{w_{\mathcal{X}_{K}},w'_{\mathcal{X}_{K}}}$.
\item\label{pcond2b}
For each $K \in \ZZ_{\ge 0}$, the sequences $\{w'_{\mathcal{Z}_K}, \ldots, w'_{\mathcal{Z}_{K+1}-1} \}$ uniformly sample $\Omega'$.
\end{enumerate} 

Based on assumptions (A)-(D), we may compute:
\begin{align}
&\frac{\mathcal{A}'_{\mathcal{Z}_K}}{\Theta} =\frac{1}{\theta N\sqrt{M}} \sum_{k=0}^{K-1}  
\sum_{m=\mathcal{Z}_{k}}^{\mathcal{Z}_{k+1}-1} \sum_{n=\mathcal{X}_{m}}^{\mathcal{X}_{m+1}-1}  e^{iS(g_n)+S(g'_n)} \notag \\
~~& \approx    \frac{1}{\theta \sqrt{M}} \sum_{k=0}^{K-1}  
\sum_{m=\mathcal{Z}_{k}}^{\mathcal{Z}_{k+1}-1}  \frac{\mathcal{X}_{m+1}-\mathcal{X}_{m}}{N} \psi(w_{\mathcal{X}_{m}})\phi(w'_{\mathcal{X}_{m}})  \label{eq:approx} \\
~~& =    \frac{1}{\theta } \sum_{k=0}^{K-1}  \left( \psi(w_{\mathcal{X}_{\mathcal{Z}_k}}) \cdot \frac{1}{\sqrt{M}} \sum_{m=\mathcal{Z}_{k}}^{\mathcal{Z}_{k+1}-1}  \frac{\mathcal{X}_{m+1}-\mathcal{X}_{m}}{N} \phi(w'_{\mathcal{X}_{m}})  \right) \notag \\
~~& =    \frac{1}{\theta } \sum_{k=1}^{K}  
\left( \psi(w_{\mathcal{X}_{\mathcal{Z}_{k}}}) \cdot \frac{1}{\sqrt{\zeta_k}}  \sum_{m'=1}^{\zeta_k}  \eta_{m',k}  \right),
\label{eq:fin}
 \end{align}
where the approximation holds for large $N$ and
\begin{equation}\label{eq:etadef}
\begin{aligned}
\zeta_k &\equiv \mathcal{Z}_{k+1}-\mathcal{Z}_k;\qquad \xi_m \equiv \mathcal{X}_{m}-\mathcal{X}_{m-1};\\
\eta_{m,k} &\equiv
\sqrt{\frac{\zeta_k}{M}} 
\left(\frac{\xi_{\mathcal{Z}_{k}+m}}{N}\right) 
\phi\left(w'_{\mathcal{X}_{\mathcal{Z}_{k}+m}}\right).
\end{aligned}
\end{equation}
Notice the similarity between (\ref{eq:first}) and (\ref{eq:fin}). Instead of a summation over $M$, there is a summation over $\zeta_k$, which has expectation $M$.  Within this summation, instead of the mean-zero i.i.d. random variables $\{\eta_k\}$, we  now have $\{\eta_{k,m}\}$ given by the complicated expression  (\ref{eq:etadef}).  By assumption, the variables $\zeta_k/{M}$ and 
$\xi_{\mathcal{Z}_{k-1}+m}/{N}$ are independent, and have expectation 1; while the additional complex factor $\phi\left(w'_{\mathcal{X}_{\mathcal{Z}_{k-1}+m}}\right)$ will vary randomly with mean zero as the process evolves.  
If we assume that $\{\eta_{k,m}\}$ are (approximately) i.i.d. mean-zero random variables, then (\ref{eq:fin}) and (\ref{eq:first}) are virtually identical, except that $\zeta_k$ in   (\ref{eq:fin}) replaces $M$ in (\ref{eq:first}).  However, $E[\zeta_k] = M$; and conditioning on the different possible values of $\zeta_k$, we may obtain the same result that the  probability density for $w_{K_{\Theta}}$ is given by $|\psi(w)|^2$.  

\begin{figure}[h]
\begin{center}
\includegraphics[width=6.5in]{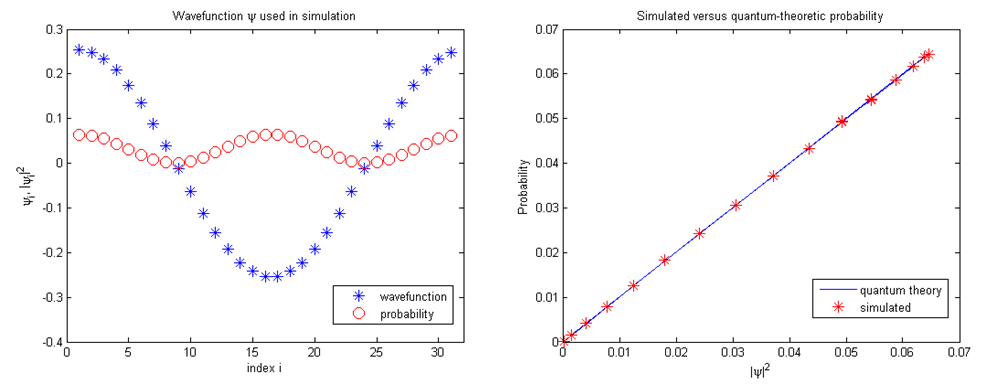}
\end{center}
\caption{(Left) Sinusoidal ``wavefunction'' used in simulation. 31 states were used with probabilities as shown. (Right) Simulation results compared to theory for $\theta=10, M=625$. Computed probabilities are based on 10 million repetitions.    \label{fig:expsim1}}
\end{figure}

\begin{figure}[h]
\begin{center}
\includegraphics[width=6.5in]{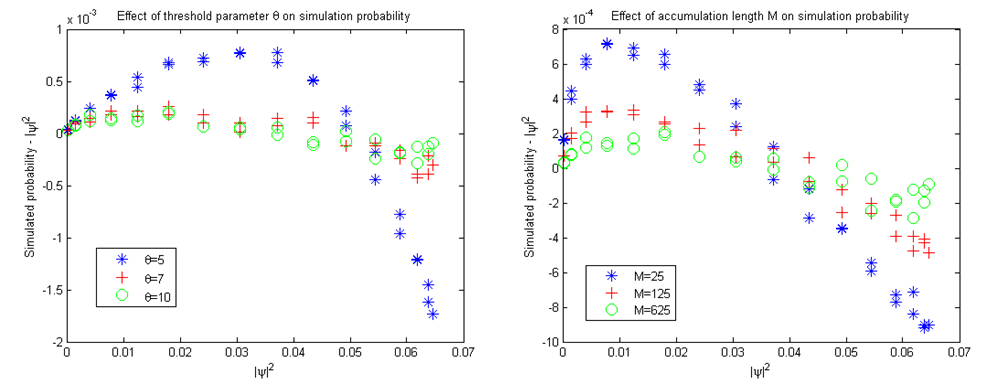}
\end{center}
\caption{Deviations of computed probabilities from quantum values for simulated adjusted accumulation model, for different values of the accumulation length $M$ and threshold parameter $\theta$ (as specified in the figure titles).  All computed probabilities are based on 10 million repetitions. \label{fig:expsim2}}
\end{figure}

Figure~\ref{fig:expsim1} shows results of simulations of the adjusted model specified in (\ref{pcond1a})-(\ref{pcond2b}).  A system with 31 discrete states was simulated, and the states' probabilities were chosen according to the sinusoidal wavefunction shown in the picture. The transition between states $w$ was determined according to a Markov chain that produced a mean dwell time of $M$, followed by a transition to one of the four nearest-neighbor states with equal probability 1/4. Parameters used were $M=625$ and $\theta=10$.  The figure shows very close agreement between quantum-theoretic probabilities and those obtained from simulation. Deviations are shown in more detail in Figure~\ref{fig:expsim2} for different values of $M$ and $\theta$. 
Small $|\psi|^2$'s are consistently overestimated, and large $|\psi|^2$'s are underestimated. 
Deviations between simulation and quantum theory decrease with increasing $M$ and $\theta$, so that the model probabilities apparently converges to quantum-theoretic values as $M, \theta \rightarrow \infty$.  

\section{Proposed Experimental Test}\label{sec:exper}
In the above model, quantum probabilities are generated by an accumulative process which essentially performs a stochastic approximation to the quantum path integrals.  In the previous section we showed that finite values of $\theta$ and $M$ introduced deviations from quantum-theoretical probabilities. In both cases, the deviations  are positive for small probabilities, but negative for large probabilities.  

Another possible source of random deviations, which we did not model in the simulation, results from the approximation
\begin{equation}
 \frac{1}{\xi_m}\sum_{n=\mathcal{X}_{m}+1}^{\mathcal{X}_{m}+ \xi_m}  e^{iS(c_n)}  \approx     \psi(w_{\mathcal{X}_{m+1}}),  \label{eq:approx2}
\end{equation}
 which was used in (\ref{eq:approx}).  If we suppose there is a random error of constant variance $\epsilon^2$ in this approximation, then by carrying through the computations it can be shown that probabilities turn out to be proportional to $|\psi(w)|^2 + \epsilon^2$ rather than $|\psi(w)|^2$.  This produces a deviation from theoretical probabilities that decreases 
linearly with increasing probability density. So the deviations from quantum-theoretic probabilities due to this effect reinforce the deviations already discussed.

 We may conclude that numerical approximation effects should introduce a deviation from quantum-theoretic probabilities that for larger probabilities decreases roughly linearly with increasing probability density.  Unfortunately, since the parameters of the process are not directly accessible, it is not possible to predict the size of the deviations. 

\section{Discussion}\label{sec:disc}
This construction provides a conceptually simple solution to many conundrums of quantum theory. 
It accounts for all quantum paradoxes, since it yields the same predictions as quantum theory (to a close approximation). It also has many advantages compared to other interpretations of quantum mechanics. It avoids the agnosticism of the Copenhagen interpretation; it circumvents the complicated branching and enormous configuration space required by the many-worlds interpretation; it requires no internal guidance system for particles as does Bohmian quantum mechanics; and it avoids difficulties with causality inherent in transactional quantum mechanics. Furthermore, unlike these other interpretations, the probabilistic nature of wavefunctions is intrinsic to the model.  

The model gives a highly non-intuitive picture of the workings of the universe. It tells us that causality is an illusion: apparent ``cause and effect'' relationships are correlations in the outcome of an  inaccessible process that occurs outside of spacetime. The Big Bang does not account for the ``origin'' of the universe, because it  is also part of the outcome of  an extra-dimensional  process which produces past, present, and future together as an entirety. (The model thus seems to imply that the universe will have finite duration.) The vacuum is not a ``boiling sea of virtual particles and antiparticles''\cite{NatGeo} as quantum field theories seem to imply,  but only appears so because of the accumulation process through which the observable universe is actualized. 

If this conceptual model of the universe proves to accurate, it has profound implications for how we may regard the world around us, and how we regard ourselves as free agents within it.  For a more thorough discussion of these implications, see  \cite{thron2021sliced}. 

\section{Acknowledgments}
Thanks to Johnny Watts for help in the preparation of this paper for publication.






\end{document}